\documentstyle[preprint,prl,aps]{revtex}
\begin{document}
\draft
\title{Streamline topology and dilute particle dynamics in a K\'arm\'an vortex street flow}
\author{ Zuo-Bing Wu } 
\address{State Key Laboratory of Nonlinear Mechanics (LNM), Institute of Mechanics, 
Academia Sinica, Beijing 100080, China}
\date{\today}
\maketitle

\begin{abstract}
Three types of streamline topology in a  K\'arm\'an vortex street flow are shown under
the variation of spatial parameters. For the motion of dilute particles
in the K\'arm\'an vortex street flow, there exist a route of bifurcation to a 
 chaotic orbit and more attractors in a bifurcation diagram for
the proportion of particle density to fluid density. Along with the increase 
of spatial parameters in the flow filed, the bifurcation process is suspended, 
as well as more and more attractors emerge.
In the motion of dilute particles, a drag term and gravity term dominate 
and result in the bifurcation phenomenon. 

\end{abstract}

\section{Introduction}

The motions of particles in a nonuniform flow have wide technological
applications, such as to forecast chemical reactions and environmental pollution.
Due to particle motion in the low Reynolds number category, the equation
of motion for a small rigid sphere in a nonuniform flow field is deduced\cite{MR}.
When the background flow is 
mainly dominated by large scale structures, the fluid viscosity is not included 
in the governing flow equation\cite{CT}.
Related studies show that even when the background flow fields are very simple,
the motions can have  abundant phenomena. 
In a periodic Stuart vortex flow, depending on the values of parameters,
the particles asymptotically concentrate
along periodic, quasiperiodic or chaotic open trajectory\cite{GL,TGL}.
In a cellular flow field, aerosol particles also
merge into isolated asymptotic trajectories, which are described by 
slow manifolds\cite{M,JM}.
Moreover, the method due to the Lagrange view of particles 
can also be applied to investigate effects of particle dispersion on
streamwise braid vortices in a plane mixing layer\cite{MM,MMR}. 

In a plane wake flow behind a circular cylinder, a regular vortex street
structure was investigated at $Re=60 - 5000$.  Along with the advance of
experimental technique, even at $Re=O(10^4)$, the regular vortex street is
obtained by using the phase-average method\cite{HH}. In particular, the phenomena
relating to organized vortex structure, such as, reconnection of vortex street\cite{Okude}
and emergence of three dimensional vortex structure\cite{W}, arouse wide interest for the
transition of a plane wake flow. Recently, particle focusing in narrow bands near
the peripheries of the vortex structures for  
the particle dispersion in a plane wake flow is observed experimentally\cite{TWCCT}. 
By considering Stokes drag, 
 the phenomenon of particle focusing is studied on two dimensional centre manifolds\cite{BDM}. 
 The regular K\'arm\'an vortex street flow
 as a model to approach the plane wake flow and investigate 
the above phenomena plays an important role.

For the motion of particles in regular vortex street flow, 
K\'arm\'an vortex spacing influences on topological
structure of background flow field. At the same time,
a density ratio as a basic parameter may have a wide range.  
 In this paper, we will consider streamline topology and dilute particle dynamics 
 in the K\'arm\'an vortex street flow in a range of density ratio.
 In Sect. II, it is shown that for the K\'arm\'an vortex street flow,
 there exist three types of global topological structure depending on the
  spatial parameters in flow field.
 Dilute particle dynamics in the K\'arm\'an vortex street flow related to
 the density ratio is investigated in Sect. III. 
 Effects of spatial parameters in flow field on dilute particle 
 dispersion is determined in Sect. IV. Finally, a brief summary is given in Sect. V. 
 
\section{Streamline topology}
\label{sec:bert}

The stream function of K\'arm\'an vortex street flow\cite{Milne} is

\begin{equation}
\Psi(x,y)=\frac{\Gamma}{4\pi} ln \frac{ch \frac{2\pi}{l}(y-h/2)- cos \frac{2\pi}{l}x}
{ch \frac{2\pi}{l}(y+h/2)+ cos \frac{2\pi}{l}x}+ \frac{\Gamma y}{2l} th \frac{\pi h}{l},
\label{eq1.1}
\end{equation}
where $\Gamma$ is the strength of vortices, $l$ and $h$ are the streamwise and
transverse spacing of vortices, respectively.
The dimensionless quantities denoted by asterisks are introduced as 
$x^*=x/l$, $y^*=y/l$, $h^*=h/l$, $u^*=u/U_\infty$, $\Gamma^*=\Gamma/(U_\infty l)$
and $\Psi^*=\Psi/(U_\infty l)$. The stream function (1) can be represented as
\begin{equation}
\Psi^*(x^*,y^*)=\frac{\Gamma^*}{4\pi} ln \frac{ch 2\pi (y^*-h^*/2)- cos 2\pi x^*}
{ch 2\pi (y^*+h^*/2)+ cos 2\pi x^*}+ \frac{\Gamma^*y^*}{2} th \pi h^*.
\label{eq1.2}
\end{equation}
From now on, the asterisks ''*'' for the dimensionless quantities in this section 
are omitted for convenience.
The stream function (2) has symmetries:
$\Psi (x+1/2,-y)=-\Psi (x,y), \Psi (x+1,y)=\Psi (x,y)$.

The associated velocity filed is given by 

\begin{eqnarray}
u_x=\frac{\partial \Psi}{\partial y}=\frac{\Gamma}{2}[\frac{sh 2\pi (y-h/2)} 
{ch 2\pi (y-h/2)- cos 2\pi x}- \frac {sh 2\pi (y+h/2)} 
{ch 2\pi (y+h/2)+ cos 2\pi x}] + \frac{\Gamma}{2} th \pi h, \nonumber \\
u_y=-\frac{\partial \Psi}{\partial x}=-\frac{\Gamma } {2} [\frac{1} 
{ch 2\pi (y-h/2)- cos 2\pi x}+\frac{1} 
{ch 2\pi (y+h/2)+ cos 2\pi x}] sin 2 \pi x,
\label{eq1.3}
\end{eqnarray}
which has singularity at the vortex centers.  
In the case of removing the singularity by the Rankine vortex\cite{WL},
 free stagnation points in velocity field (3) 
consist of centers and saddle points, which can be respectively described by
$(0,\frac{h}{2})$, $(\frac{1}{2},-\frac{h}{2})$ and
$(0,-\frac{1}{2 \pi} ln c$, $(\frac{1}{2},\frac{1}{2 \pi} ln c)$,
where $c=2/sh \pi h + sh \pi h
+ \sqrt { [2/sh \pi h + sh \pi h]^2 +1}$. 

In the symmetry $\Psi (x+1/2,-y)=-\Psi (x,y)$, it can be shown that there exists
a zero streamline between two neighboring centers $(0,\frac{h}{2})$ and $(\frac{1}{2},-\frac{h}{2})$.
Under the variation of spatial parameter $h$, we take zero streamlines to
investigate the evolution of global topological structure. The points
$y_0$, corresponding to zero streamlines passing through the $y$ coordinate, satisfy
\begin{equation}
\Psi(0,y_0)=\frac{\Gamma}{4 \pi} ln \frac{ch 2\pi (y_0-h/2)-1}
{ch 2\pi (y_0+h/2)+1} + \frac{\Gamma y_0}{2} th \pi h=0.
\label{eq1.6}
\end{equation}
Using the Newton-Raphson and bisection methods\cite{PTVF}, we solve Eq.(4) and plot
the relation $y_0 \sim h$ in Fig. 1. 
		
When $h=0$, the K\'arm\'an vortex street returns to a periodic row
of vortex pair ($u_x=\Gamma \frac{cos 2 \pi x
sh 2 \pi y} {ch^2 2 \pi y - cos^2 2 \pi x}$
and $u_y=-\Gamma \frac{sin 2 \pi x
ch 2 \pi y} {ch^2 2 \pi y - cos^2 2 \pi x}$).  
In this case, a zero streamline exists in $y_0=\infty$, 
as well as free stagnation points are only centers.
All streamlines including an embedded zero streamline for $h=0.001$, 
which approach to those for $h=0$, are drawn in Fig.~2(a).
Two neighboring vortices are divided by the zero streamline.
In Fig.~1, there exists a minimum point $(0.164,0.443434)$ marked by $o$. 
To understand this point, we 
take the derivative of $\Psi(0,y)$ with respect to $h$, as follows,

\begin{equation}
\frac{\partial \Psi(0,y)}{\partial h}
=-\frac{\Gamma}{4} [ \frac{sh 2\pi (y-h/2)}{ch 2\pi (y-h/2)-1} 
+\frac{sh 2\pi (y+h/2)}{ch 2\pi (y+h/2)+1} 
-2 \pi y /ch^2 \pi h ]=0.
\label{eq1.7}
\end{equation}
Using the Newton-Raphson and bisection methods\cite{PTVF}, we solve Eq.(5) and plot the
 solutions in Fig.~3(a). Besides a trivial solution $y=h/2$ of Eq.(5), other relations
 $\pm y \sim h $ are  symmetrical and corresponding to the maximum and minimum of
 $\Psi(0,y)$ for $h$. The point (0.164,0.443434) exists in the curve $y \sim h$ marked
 by $p$, so that $\Psi(0,0.443434)$ is a maximum of $\Psi(0,y)$ for $h$.
 We also plot the relation of $\Psi(0,0.443434)$ to
$h$ in Fig.~3(b). The maximum point $q$ in the curve confirms to the above result.
Thus, the minimum point in Fig.~1 corresponds to
the maximum of $\Psi(0,y)$ for $h$.

When $h$ increases from 0 to 0.164, $y_0$ monotonically decreases from
$\infty$ to 0.443434 in Fig.~1. All  streamlines including an embedded zero
streamline for $h=0.1$ are drawn in Fig.~2(b).  
In this case, saddle points emerge below 
or above the centers in opposition to the zero streamlines. The orbits passing through saddle
points go around the centers and come back, as well as go to other saddle points,
i.e., there exist homoclinic and heteroclinic orbits. 
 For $h=0.164$, all streamlines 
 including an embedded zero streamline are drawn in Fig.~2(c).
When $h$ increases from 0.164, $y_0$ monotonically increases from
0.443434 in Fig.~1. All streamlines  including an embedded zero  streamline
for $h=0.3$ are drawn  in Fig.~2(d). 
When $h$ increases to 0.410998, another solution of $y_0$ emerges in Fig.~1. All associated
streamlines including embedded zero  streamlines are drawn in Fig. 2(e). 
Zero streamlines, surrounding each vortex, pass through the saddle points. 
So, the zero streamlines are homoclinic and heteroclinic orbits.
When $h$ increases beyond 0.410998,
the solution becomes two ones in Fig.~1. At the time, there exist three
zero streamlines. The orbits passing through saddle
points go around the centers and to other saddle points, but do not come back, 
i.e., there exist only heteroclinic orbits.
All streamlines including three embedded zero  streamlines for $h=0.6$ are drawn in
  Fig.~2(f). Upper and lower vortices are divided by one zero streamline, as well as
  surrounded by others.

In the experiment of a plane wake flow\cite{OM}, it has been shown that 
$h$ increases downstream from
 0.2 to 0.45 at $Re=140$. We can thus conclude that three types of 
streamline topology can emerge in a plane wake flow.

\section{Dilute particle dynamics}
\label{sec:dynamics}

\subsection{Governing equations}
\label{sec:equa}

The motion of a small spherical particle in a nonuniform flow field ${\bf u}$
is governed by the momentum equation\cite{MR}

\begin{eqnarray}
\frac{\pi}{6} d^3 (\rho_P+0.5 \rho_F) \frac{d{\bf V}}{dt}= &
\frac{\pi}{6} d^3 (\rho_P- \rho_F){\bf g} +\frac{\pi}{4} d^3 \rho_F \frac{D{\bf u}}{Dt}
+3 \pi d \nu \rho_F ({\bf u}-{\bf V})f_d \nonumber\\
& + \frac{3}{2} (\pi \nu)^{1/2} d^2 \rho_F \int_0^t
\frac{1}{\sqrt{t-\tau}} (\frac{d{\bf u}}{d\tau}-\frac{d{\bf V}}{d\tau}) d\tau
+\frac{\pi}{12} d^3 \rho_F ({\bf u}-{\bf V}) \times \omega,
\label{eq2.1}
\end{eqnarray}
where ${\bf V}$ is the velocity of the particle, $d$ is the particle diameter,
$\rho$ is the density, ${\bf g}$ is the gravitational acceleration, $\nu$ is the fluid
kinematic viscosity, $\omega$ is the vorticity of the flow fluid, and the subscripts
$F$ and $P$ refer to the fluid and particle, respectively. The parameter $f_d$ 
relating to Reynolds number ($Re_P=\frac{|{\bf u-V}|d}{\nu}$) is 
described$^{\cite{CGW,TGL}}$ as

\begin{equation}
f_d=1+0.1315 Re_P^{0.82-0.05 log_{10}^{Re_P}},~~~0<Re_P <200.
\label{eq2.2}
\end{equation}
Introducing the dimensionless quantities $\delta=\rho_P / \rho_F$,
$\epsilon=1/(0.5+\delta)$, $t^*=t/T$, ${\bf V^*}={\bf V}/(l/T)$ 
and ${\bf g^*}={\bf g}/g$ ($T$ is
the particle viscous relaxation time $d^2/(18\epsilon \nu)$), we 
nondimensionlize Eq. (6) and ignore the Basset history term. Thus,
Eq. (6) can be described by
\begin{equation}
\frac{d{\bf V}}{dt}=B{\bf g}+\frac{3}{2}\epsilon A^2 {\bf u} \bullet \nabla {\bf u}
+(A{\bf u-V})f_d +\frac{1}{2} \epsilon A(A{\bf u-V}) \times \omega.
\label{eq2.25}
\end{equation}
where $A=U_\infty T/l$, $B=(1-1.5\epsilon) T^2 \frac{g}{l}$
and  the asterisks ''*'' for the dimensionless quantities in this section 
are omitted for convenience.
Moreover, the Reynolds number is written as $Re_P=\overline{Re}_P/A|A{\bf u-V}|$ 
($\overline{Re}_P=U_\infty d/ \nu$).

The flow field ${\bf u}$ is chosen to be the K\'arm\'an vortex street flow (3), 
where a parameter $k$ is introduced to remove the singularities.
The modified stream function (2) is described as
\begin{equation}
\Psi(x,y)=\frac{\Gamma}{4\pi} ln \frac{ch 2\pi(y-h/2)-k cos 2\pi x}
{ch 2\pi(y+h/2)+k cos 2\pi x}+\frac{\Gamma y}{2} \frac{sh 2\pi h} {ch 2\pi h +k}.
\label{eq2.3}
\end{equation}
 The corresponding velocity field is given by
\begin{eqnarray}
u_x=\frac{\partial \Psi}{\partial y}=\frac{\Gamma}{2}[\frac{sh 2\pi(y-h/2)} 
{ch 2\pi(y-h/2)-k cos 2\pi x}- \frac {sh 2\pi(y+h/2)} 
{ch 2\pi(y+h/2)+k cos 2\pi x}] + \frac{\Gamma}{2} \frac{sh 2\pi h} {ch 2\pi h +k},\nonumber\\
u_y=-\frac{\partial \Psi}{\partial x}=-\frac{\Gamma k} {2} [\frac{1} 
{ch 2\pi(y-h/2)-k cos 2\pi x}+\frac{1} 
{ch 2\pi(y+h/2)+k cos 2\pi x}] sin 2\pi x.
\label{eq2.4}
\end{eqnarray}
Only  when $k=1$, the velocity field satisfies the Euler equation.
Since the error increases with deviation of $k$ from 1, we take $k=0.99$ as
an approximation in this simulation.

The particle motion is described by a four-dimensional, nonlinear autonomous dynamical
system of the form
\begin{eqnarray}
\dot{x}=V_x,\nonumber\\
\dot{y}=V_y,\nonumber\\
\dot{V_x}=\frac{3}{2} \epsilon A^2 {\bf u} \bullet \nabla u_x +(Au_x-V_x)f_d 
	+ \frac{1}{2} \epsilon A \omega (Au_y-V_y),\nonumber\\
\dot{V_y}=\frac{3}{2} \epsilon A^2 {\bf u} \bullet \nabla u_y +(Au_y-V_y)f_d 
	- \frac{1}{2} \epsilon A \omega (Au_x-V_x)-B.
\label{eq2.5}
\end{eqnarray}

The parameters in Eqs. (11) can be taken as $U_\infty=4 m/s$ 
and $d=5 \times 10^{-5} m$ from \cite{TWCCT}. Since air is chosen as the fluid media 
in the flow, the properties of fluid in Eqs. (11) are described as 
$\rho=1.225 kg/m^3$  and $\nu=1.45 \times 10^{-5} m^2/s$\cite{Panton}.
Moreover, the parameters concerning the K\'arm\'an vortex street flow are chosen as
$l=0.1 m$, $h=0.3$ and $\Gamma=1$. In the following, we investigate the motion 
of particles under the variation of density ratio $\delta$.
Initial values of particles are taken as points in the flow field
and their corresponding flow filed velocity. Using a fourth-order Runge-Kutta 
algorithm, we integrate Eqs. (11) with a time size $\tau=0.001$. After discarding 
transients, we plot points at $x=0$ as a bifurcation diagram. 
When the time size is changed to $\tau=0.01$, the bifurcation diagram  
is still preserved.

In order to understand the dilute particle dynamics, we analyze
orders of magnitude of parameters in Eq.~(8). 
For giving physical values in the calculation,
the parameter $f_d$ appear to be on the order 1.
 When $\delta$ is taken as the order $10^3 - 10^4$ and $d$ is fixed as the order $10^{-5}m$, 
the parameters $\epsilon$, $T$, $A$ and $B$ appear 
to be on the order $10^{-3} - 10^{-4}$, $10^{-3} s - 10^{-2} s$, $10^{-2} - 10^{-1}$ 
and $10^{-4} - 10^{-2}$, respectively.  In this case, the stress tensor term of fluid 
$\frac{3}{2} \epsilon A^2 {\bf u} \bullet \nabla {\bf u}$ 
and lift force term $\frac{1}{2} \epsilon A(A{\bf u-V}) \times \omega$
have smaller orders than the drag term $(A{\bf u}-{\bf V})f_d$ 
and gravity term $B{\bf g}$ in the Eq.~(8). Therefore,
the Eq.~(8) is dominated by the drag term and gravity term . 

Moreover, concerning the stability of particle orbits, we take only 
the drag and gravity terms in Eq.~(8) to obtain an approximate 
fundamental matrix $U^t_{\bf x}$. By using the straightforward technique\cite{SN,WSSV},
a complete Lyapunov spectrum is determined as follows
\begin{equation}
\lambda=\lim_{t \rightarrow \infty} \frac{1}{t} log \frac{\parallel U_{\bf x_0}^{t} {\bf e_1^0}
\wedge U_{\bf x_0}^{t} {\bf e_2^0}  \wedge U_{\bf x_0}^{t} {\bf e_3^0} 
\wedge U_{\bf x_0}^{t} {\bf e_4^0} \parallel} {\parallel {\bf e_1^0} \wedge {\bf e_2^0} \wedge 
{\bf e_3^0} \wedge {\bf e_4^0} \parallel}
=\lim_{n \rightarrow \infty} \frac{1}{n \tau} \Sigma_{i=0}^{n-1} 
log \frac{\parallel U_{\bf x_i}^{\tau} {\bf e_1^i}
\wedge U_{\bf x_i}^{\tau} {\bf e_2^i}  \wedge U_{\bf x_i}^{\tau} {\bf e_3^i} 
\wedge U_{\bf x_i}^{\tau} {\bf e_4^i} \parallel} {\parallel {\bf e_1^i} \wedge {\bf e_2^i} \wedge 
{\bf e_3^i} \wedge {\bf e_4^i} \parallel},
\label{eq2.6}
\end{equation}
where $\wedge$ and $\parallel \circ \parallel$ are an exterior product and a norm
with respect to some Riemannian metric, respectively. After each time integration,
the set of bases  $\{\bf e_1^i, e_2^i, e_3^i, e_3^i\}$ is exchanged by 
using the Gram-Schmidt reorthonormalization procedure. In the calculation
of the complete Lyapunov spectrum, the time size is taken as $\tau=0.001$.

\subsection{Numerical results}
\label{sec:bif}
  
In Fig.~4, along with the increase of density ratio, a bifurcation diagram 
of $y$ versus $\delta$ is drawn. When $\delta < 6.5 \times 10^2$, the velocity
of particle dispersion is very slow, so most of the particle trajectories
are preserved near the street. When $\delta \geq 6.5 \times 10^2$, 
besides the particle dispersion, global trajectories
of particles converge to two attractors: one  above the
 street, the other in the street. 
 For the attractor above the street in $\delta=6.5 \times 10^2 - 4.4 \times 10^3$,
 it evolves as a period-1 orbit. At $\delta=4.4 \times 10^3$, 
 the  period-1 orbit
bifurcates to a  period-2 orbit. In $\delta=4.4 \times 10^3 - 7.0 \times 10^3$,
the attractor evolves as a period-2 orbit. At $\delta=7.0 \times 10^3$, 
the  period-2 orbit bifurcates to a  period-4 orbit.
In $\delta=7.0 \times 10^3 - 7.9 \times 10^3$, the attractor evolves as a period-4 orbit.
In $\delta=7.9 \times 10^3  - 1.0 \times 10^4$, the
 period-4 orbit bifurcates to a period-8 orbit and further to a  quasi-periodic or chaotic orbit.
At $\delta=1.0 \times 10^4$, a crisis happens, so that the  
quasi-periodic or chaotic orbit disappears. Moreover, at $\delta=7.3 \times 10^3$, 
 a period-3 orbit emerges as another attractor above the street. 
 In $\delta=7.3  \times 10^3 - 7.9 \times 10^3$, 
 the period-3 orbit bifurcates further to a  quasi-periodic or chaotic orbit.
At $\delta= 7.9 \times 10^3$, a crisis happens, so that the  quasi-periodic 
or chaotic orbit disappears. 
The bifurcation procedure of period-3 orbit differs from that for the three
dimensional Lorenz equation\cite{Lorenz}.
For the attractor in the street, a  period-1 orbit
emerges at $\delta=3.3 \times 10^3$. It preserves 
in $\delta=3.3 \times 10^3 - 1.2 \times 10^4$.

 In Fig.~5, we plot some typical examples for different values of $\delta$. 
 A  period-1 orbit above the street for $\delta = 1.0 \times 10^3$, as well as 
 all corresponding streamlines are drawn  in Fig.~5(a). On the period-1 orbit,
 the particles move from left to right. In Fig.~6, for the period-1 orbit,
 the maximal Lyapunov exponent is -0.230, so the orbit is stable.
In order to display the  dispersion of particles in the flow field, we also plot
the basin of attraction in Fig.~7(a). 
In the motion of particles, the points
corresponding to those in the basin of attraction suspend on the period-1 orbit.
At the same time, the points corresponding to those outside the basin of attraction 
escape from the central region of flow. 
 For $\delta = 3.5 \times 10^3$, in Fig.~5(b),
two period-1 orbits distribute above and in the street, respectively. 
On the period-1 orbit above the street, the particles move from left to right.
In Fig.~6, for the period-1 orbit,
 the maximal Lyapunov exponent is -0.254, so the orbit is stable.
 But, on the orbit in the street, the particles move in an opposite direction,
 i.e., from right to left. The maximal Lyapunov exponent is -1.144, so the orbit is stable.
 For the two period-1 orbits, the corresponding basins of attraction are plotted 
in Fig.~7(b). One is similar to that in Fig.~7(a), the other is distributed 
in two local zones. The basin for the orbit in street is surrounded by that
for the orbit above street. 
For $\delta = 6.0 \times 10^3$, a period-2 orbit above the street and 
a period-1 orbit in the street are drawn in Fig.~5(c). 
On the period-2 orbit above the street, the particles move from left to right.
In Fig.~6, for the period-2 orbit,
 the maximal Lyapunov exponent is -0.604, so the orbit is stable.
 But, on the orbit in the street, the particles move in an opposite direction,
 i.e., from right to left. The maximal Lyapunov exponent is -1.268, so the orbit is stable.
 The corresponding basin of attraction is plotted in Fig.~7(c), which is
 similar to Fig.~7(b).  For $\delta = 7.5 \times 10^3$, 
 a combined period-4 orbit with a period-3 above the street and a period-1 orbit
in the street are drawn in Fig.~5(d).
On the period-4 and period-3 orbits above the street, the particles move from left to right.
But, on the orbit in the street, the particles move in an opposite direction,
 i.e., from right to left. In Fig.~6, for the period-4, period-3 and period-1 orbits,
 the maximal Lyapunov exponent are -1.003, -0.245 and -1.356, respectively, 
 so the orbits are stable.
  The corresponding basin of attraction is plotted in Fig.~7(d). 
 The geometry of basin is different from above ones in Figs.~7(a)-(c).
 Some points near $y=1$ escape from the basin of attraction in the dispersion of particles.
 For $\delta = 9.5 \times 10^3$, a quasi-periodic or chaotic orbit above the street
 and a period-1 orbit in the street are drawn in Fig.~5(e).
 On the orbit above the street, the particles move from left to right.
 In Fig.~6, for the orbit,
 the maximal Lyapunov exponent is 0.850, so the orbit is chaotic.
 But, on the orbit in the street, the particles move in an opposite direction,
 i.e., from right to left. The maximal Lyapunov exponent is -1.356, so the orbit is stable.
  The corresponding basin of attraction is plotted in Fig.~7(e). 
  For the period-1 orbit in the street, the geometry of basin is similar to that
  in Fig.~7(d). But, for the chaotic orbit above the street, 
  escaped points in Fig. 7(e) permeate into the basin of attraction in Fig. 7(d),
  so that the geometry of basin is fractal.
 From those examples, we can conclude that along with the increases of density ratio,
 initial points distributed in the central region of flow escape more and more.
 At the same time, the particle trajectories bifurcate from periodic orbits to
 chaotic orbits.

 In order to explain the existence of attractors in the flow field,
 we take the period-1 and period-2 orbits for $\delta=6.0 \times 10^3$ as an example. 
 In Fig.~8, we present the values of $V_x$, $V_y$, $u_y$ and
 $(Au_y-V_y)f_d-B$ along two orbits. For the period-2 orbit above the street,
 we plot it in two times of the streamwise periodic length in Fig.~8(a).
 Since $V_x>0$ in $x \in [0,2)$, the motional direction of particles is 
 from left to right. (1) When a particle moves from $x=0$ to $x=0.486$, the negative term
 $(Au_y-V_y)f_d-B$ causes decrease of $V_y$ from 0.050 to -0.285.
 At the same time, it leads to increase the term $(Au_y-V_y)f_d-B$. (2)
 When the particle moves from $x=0.486$ to $x=0.930$, the positive term
 $(Au_y-V_y)f_d-B$ causes increase of $V_y$ from -0.285 to 0.404.
 At the same time, it leads to decrease the term $(Au_y-V_y)f_d-B$. 
 (3) When the particle moves from $x=0.930$ to $x=1.554$, 
 the negative term $(Au_y-V_y)f_d-B$ causes decrease of $V_y$ from 0.404 to -0.115.
 At the same time, it leads to increase the term $(Au_y-V_y)f_d-B$.
 (4) When the particle moves from $x=1.554$ to $x=1.946$, 
 the positive term $(Au_y-V_y)f_d-B$ causes increase of $V_y$ from -0.115 to 0.067.
 At the same time, it leads to decrease the term $(Au_y-V_y)f_d-B$.
 (5) When the particle moves from $x=1.946$ to $x=2$, 
 the negative term $(Au_y-V_y)f_d-B$ causes decrease of $V_y$ from 0.067 to 0.050.
 At the same time, it leads to increase the term $(Au_y-V_y)f_d-B$.
 According to the periodic boundary condition, when the particle reaches $x=2$,
 it goes back to $x=0$. Therefore, the combination of the 
drag term and gravity term in the vertical direction has a periodic vibration
along with the variation of the vertical velocity of particles and 
makes the period-2 orbit. 
In Fig.~8(b), for the period-1 orbit in the street,
we plot it in the streamwise periodic length.
Since $V_x<0$ in $x \in [0,1)$, the motional direction of particles is 
 from right to left. In a similar way to Fig.~8(a), the term $(Au_y-V_y)f_d-B$
 brings into a periodic vibration of particles and makes the period-1 orbit.  
In Ref.\cite{BDM}, by considering the drag term, 
the essential dynamics takes place on the two-dimensional centre manifolds. 
To compare with the result, we
 find the bifurcation process disappears when eliminate the gravity term in Eq.~(8). 
From the above observation,
we can conclude that the drag term and gravity term lead to the bifurcation behavior 
in dilute particle dispersion. The motional direction of particles 
and distribution of $u_y$
determine the vertical position of attractors.
 
\section{Effects of spatial parameters in flow field on dilute particle dispersion}
\label{sec:spat}

In the Sect. II, we show three types of streamline topology in the
K\'arm\'an vortex street flow at different spatial parameter $h$.
In the Sect. III, we present the dilute particle dynamics for the flow field
with $h=0.3$. In the following, we increase the
spatial parameter $h$ to consider its effects on dilute particle dispersion.

For $h=0.410998$, along with the increase of density ratio,
a bifurcation diagram $y$ versus $\delta$ is drawn in Fig.~9. The global evolution
  is similar to that in Fig.~4. 
However, the bifurcation of period-1 attractor above the street is delayed.
The period-1 orbit above the street bifurcates to a period-2 orbit
at $\delta = 6.6 \times 10^3$.
The period-2 orbit bifurcates to a period-4 orbit
at $\delta = 1.04 \times 10^4$. 
In $\delta=1.04 \times 10^4 - 1.2 \times 10^4$, the period-4 orbit  
bifurcates further to a quasi-periodic or chaotic orbit.
A period-3 orbit above the street emerges at $\delta = 7.0 \times 10^3$ and 
 bifurcates in $\delta=7.0 \times 10^3 - 8.3 \times 10^3$. 
 At  $\delta = 8.3 \times 10^3$, a crises happens, so that the 
quasi-periodic or chaotic orbit above the street disappears. 
Moreover, the occurrence of period-1 attractor in the street
is shifted earlier at $\delta = 2.0 \times 10^3$. A bifurcation of
period-3 orbit in the street emerges in $\delta = 9.4 \times 10^3 - 1.1 \times 10^4$.

For $h=0.6$, along with the increase of density ratio, 
a bifurcation diagram $y$ versus $\delta$ is drawn in Fig.~10. The global evolution
 is similar to that in Fig.~9. However, the bifurcation of period-1 attractor 
 above the street is delayed. The period-1 orbit above the street bifurcate 
 to a period-2 orbit at $\delta = 9.1 \times 10^3$. 
 The period-2 orbit is preserved in $\delta = 9.1 \times 10^3 - 1.2 \times 10^4$.  
 At the same time, the occurrence of period-1 attractor 
 in the street is shifted earlier at $\delta = 2.0 \times 10^2$. Moreover, 
 at $\delta = 7.6 \times 10^3$, a period-4 and a period-3 orbits emerge above the street. 
 They bifurcate further to two quasi-periodic or chaotic orbits and
 disappear at $\delta= 8.4 \times 10^3$ and $\delta= 8.5 \times 10^3$, respectively.
 In $\delta = 8.0 \times 10^3 - 9.4 \times 10^3$,
 a similar bifurcation process of more attractors emerges in the street.
 In $\delta = 8.0 \times 10^3 - 8.8 \times 10^3$, a period-4 orbit preserves
 in the street. In $\delta = 8.4 \times 10^3 - 9.4 \times 10^3$, 
 a bifurcation of period-3 orbit exists in the street.

Thus, along with the increase of $h$, a period-1 orbit above the street 
bifurcates slower to a quasi-periodic or chaotic orbit, as well
as a period-1 orbit in the street emerges earlier. 
Besides the main attractors in the flow field, more and more 
local attractors also appear above and in the street.

\section{Conclusion}
\label{sec:sum}
 Under the variation of spatial parameters, we have shown 
 three types of streamline topology in a K\'arm\'an vortex street flow.
 For the motion of dilute particles in the K\'arm\'an vortex street flow, 
 there exist a route of bifurcation to a chaotic orbit  and more
 attractors in a bifurcation diagram for the proportion of particle density
 to fluid density. Along with the increase of spatial parameters in the flow filed, 
 the bifurcation process is suspended, as well as more and more attractors emerge.
 In the motion of dilute particles, a drag term and gravity term dominate 
 and result in the bifurcation phenomenon.

\acknowledgments 
This work was supported in part by the National Key Program 
for Developing Basic Science G1999032801-11.


\figure{Fig. 1 Points of zero streamlines in the y coordinate.}
\figure{Fig. 2 All streamlines including corresponding embedded zero streamlines 
with centers marked by dots and saddle points marked by crosses
for (a) $h=0.001$;(b) $h=0.1$;(c) $h=0.164$;
(d) $h=0.3$; (e) $h=0.410998$; (f) $h=0.6$.}
\figure{Fig. 3 (a)A relation $y \sim h$ for $\Psi_{max/min}(0,y)$;
 (b) Stream function relating to $h$ at (0,0.443434).}
\figure{Fig. 4 A bifurcation diagram with $h=0.3$ for a continuous range of $\delta$ showing
the vertical position of particles at $x=0$. The dense bifurcation zone 
$\delta \in [7000,10000]$ above the street is enlarged and
redrawn in the bottom-right corner of the figure.}
\figure{Fig. 5 Typical trajectories in Fig.~4 for (a) $\delta=1.0 \times 10^3$ and all
streamlines; (b)  $ \delta=3.5 \times 10^3$;
(c) $\delta=6.0 \times 10^3$; (d) $\delta=7.5 \times 10^3$; (e) $\delta=9.5 \times 10^3$.}
\figure{Fig. 6 Variation of the maximal Lyapunov exponent $\lambda_{max}$ with $\delta$.
 Besides the period-3 orbit for $\delta=7.5 \times 10^3$ labeled by a triangle, 
 the attractors above and in the street are marked by dots and crosses, respectively.}
\figure{Fig. 7 Basins of attraction corresponding to the typical trajectories
in Fig. 5 for (a) $\delta=1.0 \times 10^3$; (b)  $ \delta=3.5 \times 10^3$;
(c) $\delta=6.0 \times 10^3$; (d) $\delta=7.5 \times 10^3$; (e) $\delta=9.5 \times 10^3$.
In Figs.~7(b)-(e), besides two local zones marked by thinner lines 
for the basin of period-1 orbit in the street, the global zone presents 
the basin for attractors above the street.}
\figure{Fig. 8 Distribution of $V_x$, $V_y$,  $u_y$ and $(Au_y-V_y)f_d-B$ along $x$
 on (a) a period-2 orbit above the street; 
 (b) a period-1 orbit in the street for $\delta=6.0 \times 10^3$. } 
\figure{Fig. 9 A bifurcation diagram  with $h=0.410998$ for a continuous range of 
$\delta$ showing the vertical position of particles at $x=0$. The dense bifurcation zone 
$\delta \in [7000,8300]$ above the street is enlarged and
redrawn in the bottom-right corner of the figure.}
\figure{Fig. 10 A bifurcation diagram  with $h=0.6$ for a continuous range of $\delta$ showing
the vertical position of particles at $x=0$. The dense bifurcation zone 
$\delta \in [7600,8500]$ above the street is enlarged and
redrawn in the bottom-right corner of the figure.}

\end{document}